# Aluminum Nanoparticles with Hot Spots for Plasmon-Induced Circular Dichroism of Chiral Molecules in the UV Spectral Interval

*Lucas V. Besteiro,[*] Hui Zhang, Jérôme Plain, Gil Markovich, Zhiming Wang, Alexander O. Govorov[*]*

Dr. Lucas V. Besteiro
Department of Physics and Astronomy, Ohio University, Athens, Ohio 45701, USA
E-mail: vazquezb@ohio.edu

Dr. Hui Zhang
Department of Electrical and Computer Engineering, Laboratory for Nanophotonics, Rice University, Houston, Texas 77005, USA

Prof. Jérôme Plain, Institut Charles Delaunay, LNIO, Université de Technologie de Troyes, CNRS UMR 6281, Troyes Cedex CS 42060-10004, France

Prof. Gil Markovich
School of Chemistry, Raymond and Beverly Sackler Faculty of Exact Sciences, Tel Aviv University, Tel Aviv 6997801, Israel

Prof. Zhiming Wang
Institute of Fundamental and Frontier Sciences, University of Electronic Science and Technology of China, Chengdu 610054, PR China; State Key Laboratory of Electronic Thin Films and Integrated Devices, University of Electronic Science and Technology of China, Chengdu 610054, PR China

Prof. Alexander O. Govorov
Department of Physics and Astronomy, Ohio University, Athens, Ohio 45701, USA
E-mail: govorov@ohio.edu

**Abstract**

Plasmonic nanocrystals with hot spots are able to localize optical energy in small spaces. In such physical systems, near-field interactions between molecules and plasmons can become especially strong. This paper considers the case of a nanoparticle dimer and a chiral biomolecule. In our model, a chiral molecule is placed in the gap between two plasmonic nanoparticles, where the electromagnetic hot spot occurs. Since many important biomolecules have optical transitions in the UV, we consider the case of Aluminum nanoparticles, as they offer strong electromagnetic enhancements in the blue and UV spectral intervals. Our calculations show that the complex composed of a chiral molecule and an Al-dimer exhibits strong CD signals in the plasmonic spectral region. In contrast to the standard Au- and Ag-nanocrystals, the Al system may have a much better spectral overlap between the typical biomolecule's optical transitions and the nanocrystals' plasmonic band. Overall, we found that Al nanocrystals used as CD antennas exhibit unique properties as compared to other commonly studied plasmonic and dielectric materials. The plasmonic systems investigated in this study can be potentially used for sensing chirality of biomolecules, which is of interest in applications such as drug development.

# 1. Introduction

Circular dichroism (CD) is a powerful tool to study structural properties of molecules[1,2] and nanocrystals (NCs).[3] Plasmonic NCs are especially interesting since they can strongly absorb and scatter visible light and, furthermore, plasmonic NCs are known for their strong local enhancement of electromagnetic fields. When a chiral biomolecule is attached to a nonchiral metal NC, the chiral response of a biomolecule becomes transferred to the plasmonic NC via near-field dipolar and multi-polar interactions.[4,5] Circular dichroism spectroscopy, which quantifies the difference between the assembly's responses to left and right circularly polarized light, is a well-established experimental method to measure the chiral properties of an assembly. In the biomolecule-NC assembly, CD spectra typically show modified molecular lines, in addition to including new lines coming from plasmon excitations in the NCs. It was demonstrated in a large number of experiments[3,6,7] that it is possible to observe plasmon-induced CD in the visible spectral interval using NCs made of gold and silver. It is a very interesting possibility, since the original CD signals of important biomolecules (such as proteins and DNA) are typically in the UV. In addition to the material composition of an assembly, its design also plays an important role. For example, the plasmon-induced CD can be strongly enhanced by using NC assemblies including hot spots with strongly enhanced electromagnetic fields. This behavior of bio-assemblies was proposed[5,8] and then observed experimentally.[9,10] We should also note that such plasmonic hot spots play an important role in several other optical and energy-related phenomena, including Surface Enhanced Raman Spectroscopy (SERS),[11,12] plasmon-enhanced fluorescence,[13-15] hot-electron generation,[16-20] nonlinear optics[21] and photo-thermal effect.[22] As a review on the science and technology relating to plasmonic hot spots, one can recommend ref. [23].

The intensities and shapes of plasmon-induced CD spectra depend strongly on the resonant conditions for the plasmon bands and the molecular transitions. In the bio-plasmonic assemblies based on gold and silver, the plasmonic and molecular absorption lines are typically off resonances,[4,8] since the lines for important biomolecules (DNA, proteins, etc.) are typically in the UV range and the plasmonic resonances are in the visible. This, of course, creates certain limitations for the enhancement factors of the CD signals.[5] Therefore, it is interesting to look at the other available plasmonic materials. One interesting possibility is to introduce aluminum NCs,[24-26] which have strong plasmonic resonances in the UV. In this case, one can expect both new chiro-optical properties and stronger enhancement effects. The field of Al NPs is developing rapidly by utilizing specialized fabrication technologies suitable for aluminum, which is highly reactive. The fabrication of Al NPs involves lithography,[26] nanosphere lithography[27,28] and colloidal synthesis.[29,30] Colloidal suspensions of Al-NCs are conveniently realized by releasing nanocrystals from a substrate into a solution.[31]

Here we study theoretically the resulting CD signal of bio-assemblies composed of chiral biomolecules and Al-NCs **(Figure 1)**. To accomplish this study, we make use of the quantum formalism previously developed by us in refs. [4,5,8]. The models that we develop here consist of one or two Al-NCs and a single molecular dipole (Figure 1). For comparison purposes, alongside the Al-NCs, we introduce NCs made of other typical materials, such as $SiO_2$, Si, Au and Ag (Figure 1). From our calculations, we see that aluminum NCs used as antennas for chiral biomolecules indeed exhibit very unique properties. Additionally to what was reported before in refs. [4,5,8], we notice striking differences between the CD responses from bio-assemblies based on semiconductor and metal NCs.

At this moment, there is an extended literature on the plasmonic CD effects using gold and silver NCs.[3,6] Regarding the CD effects using aluminum NCs, we are only aware of one publication, ref. [26]. This paper[26] reported relatively large Al-NCs fabricated by lithography. For such NCs, the effects of chiral molecules on the CD spectra of Al-NCs are of moderate strength since relatively large NCs with smooth shapes typically have a limited plasmonic near-field enhancement. In contrast to ref. [26], we here propose the usage of Al-NC dimers with small sizes, which have strong plasmonic hot spots under illumination. For the case of Ag-Ag nanoantennas, such plasmonic dimers did indeed demonstrate anomalously strong plasmon-induced CD signals,[3,9] as it was predicted by the theory.[5] For small Al-based nanoantenna systems, our calculations show large enhancement factors for the plasmonic CD signals. We also observe in our calculations novel properties of plasmon-induced CD as compared to the cases of gold and silver. In particular, our calculations show that the plasmon CD bands in the Al-based systems are inverted as compared to the Au- and Ag-based plasmonic complexes. This supports our theory and understanding of the plasmon-induced CD previously developed in refs. [4,5,8].

## 2. Formalism

In this paper we follow the formalism developed by us in ref. [5], where one can find many detailed derivations. Here we briefly describe the main points of the modeling procedure. To describe the hybrid bio-plasmonic complexes (Figure 1) composed of a chiral dipole and crystalline NCs, we use a convenient hybrid approach[4,5,8] that involves both quantum and classical theoretical methods, as described below. In our approach, the NCs are classical objects, whereas the chiral molecule is approximated as a point-like two-level system. This molecular system has two quantum states and, under illumination, it acquires nonzero optical electric and magnetic moments. (Figure 1a). These transition moments will be given by the corresponding matrix elements for the molecular quantum states, denoted as $\boldsymbol{\mu}_{12}$ and $\mathbf{m}_{21}$. The electric dipole moment of a molecule ($\boldsymbol{\mu}_{12}$) can be either normal or parallel to the NC surface (Figure 1a). This is important for the formation of the plasmon-induced CD that originates from the interaction between the dipole and the plasmon component. The direction of the magnetic dipole moment ($\mathbf{m}_{21}$) with respect to the vector $\boldsymbol{\mu}_{12}$ is also essential, since their relative orientation controls the magnitude and sign of the whole CD band. Optical responses of the crystalline NCs were computed using empirical local dielectric functions: ref. [32] for Al, ref. [33] for Ag, and ref. [34] for Au. In the past, our approach[4,8] was applied to a variety of bio-plasmonic systems and represented the basis to understand many experiments.

The formalism[5] was developed for small complexes, which can be described using plasmonic near-fields. The near-fields were calculated within the quasistatic approximation by using the multipole expansion method.[5] The dynamic electric fields of individual NCs were expressed in the form of sums over spherical harmonics. Then, the fields were calculated numerically by using a sufficiently large number of harmonics, $1 \leq L \leq 40$, where $L$ is the angular momentum of a harmonic. The near-fields were computed as a response on the external field and on the oscillating dipole of the molecule. We also note that our formalism is based on the quantum treatment of a single molecular dipole interacting with metal NCs. Our formalism allows us to understand the fundamental optical properties of colloidal assemblies which incorporate chiral and plasmonic elements. Another approach, which can also be applied to the available experimental systems, is based on the concept of a chiral medium.[35,36]

## 2.1. Light-matter interaction

Chiral biomolecules are quantum objects and the interaction of light with a molecule should be described by a quantum operator,

$$\hat{V} = -\hat{\boldsymbol{\mu}} \cdot \mathbf{E}_{mol} - \hat{\mathbf{m}} \cdot \mathbf{B}_{mol} = \hat{V}_\omega \cdot e^{-i\omega t} + \hat{V}_\omega^\dagger \cdot e^{i\omega t}, \quad (1)$$

where the operators of electric and magnetic dipole moments are given by

$$\hat{\boldsymbol{\mu}} = -|e|\mathbf{r}, \quad \hat{\mathbf{m}} = -\frac{|e|}{2mc_0}[\mathbf{r} \times \hat{\mathbf{p}}],$$

respectively. The interaction of the molecule with the electromagnetic fields (Equation 1) is computed with the total local fields acting on the molecule: $\mathbf{E}_{mol}$ and $\mathbf{B}_{mol}$. The plasmonic NCs strongly alter the electric fields inside the complex and the resulting electric field should be written as $\mathbf{E}_{mol} = \mathbf{E}_0 + \mathbf{E}_{ind}$, where $\mathbf{E}_0$ and $\mathbf{E}_{ind}$ are the external and plasmonically-induced contributions, respectively. The external field is a planewave and it has an amplitude, polarization and wave vector, $\mathbf{E}_0 = E_0 \mathbf{e}_0 e^{i\mathbf{k}\mathbf{r}}$.

The steady state of a molecule under illumination is described with the equation of motion:

$$\hbar \frac{\partial \rho_{nm}}{\partial t} = i \langle n |[\hat{\rho}, \hat{H}_0 + \hat{V}]| m \rangle - (\hat{\Gamma} \cdot \rho)_{nm}, \quad (2)$$

where $\hat{\rho}$ is the density matrix operator, $\hat{H}_0$ and $|m\rangle$ are the Hamiltonian and the quantum states of a molecule, respectively; $\hat{\Gamma}$ is the relaxation operator described in refs. [8] and [37]. Our model approximates the molecule as a two-level system and, therefore, we involve only two quantum states, $|1\rangle$ and $|2\rangle$ (**Figure 3**a). Correspondingly, these states have energies $E_1$ and $E_2$, and the optical transition energy is given by $\hbar\omega_0 = E_2 - E_1$. The derivation and solution of Equation 2 can be found in refs. [5,8]. Here we simply write down the result from ref. [8]:

$$\sigma_{22} = \frac{1}{\Gamma_{22}} \frac{2\Gamma_{12} \cdot |\langle 2|V_\omega|1\rangle|^2}{|\hbar\omega - \hbar\omega_0 + i\Gamma_{12}|^2},$$

$$\hat{V}_{\omega,\text{mol}} = -\hat{\boldsymbol{\mu}} \cdot \mathbf{E}_{\omega,\text{mol}} - \hat{\mathbf{m}} \cdot \mathbf{B}_{\omega,\text{mol}},$$

where $\Gamma_{12}$ is the internal broadening of the molecular transition. Then, the dissipation inside the molecule has the form:

$$Q_{\text{molecule}} = \hbar\omega_0 \cdot \frac{\sigma_{22} \cdot \Gamma_{22}}{\hbar} = \omega_0 \cdot \sigma_{22} \cdot \Gamma_{22}, \tag{3}$$

where $\Gamma_{22} = \hbar/\tau_{\text{mol}}$ describes the energy relaxation and $\tau_{\text{mol}}$ is the excitonic lifetime. Now we look at the dissipation inside the plasmonic components. Since plasmonic NCs are of relatively large sizes, a classical approach to the plasmonic NC is appropriate. Therefore, we use the convenience of the local dielectric function of the metal and write down the dissipation inside the metal NCs in the standard way:

$$Q_{NC} = \int_{\text{metal}} dV \langle \mathbf{j} \cdot \mathbf{E} \rangle_t = \text{Im}(\varepsilon_{NC}) \frac{\omega}{2\pi} \int_{\text{metal}} \mathbf{E}_\omega^{in} \mathbf{E}_\omega^{in*} \tag{4}$$

where **j** is the electric current density and $\varepsilon_{NC}$ is the local dielectric constant. The current density and the electric field are related via the standard equation $\mathbf{j}_\omega = -i\omega(\varepsilon_{NC} - 1)/4\pi \cdot \mathbf{E}_\omega^{in}$. The electric field inside a NC should be calculated in the following way:

$$\mathbf{E}_\omega^{in} = \mathbf{E}_{\omega, NC+external} + \mathbf{E}_{\omega, dipole},$$

where $\mathbf{E}_{\omega, NC+external}$ is the field induced inside the plasmonic NCs in the absence of a molecule and $\mathbf{E}_{\omega, dipole}$ is the field created by the optically-driven molecular dipole. We should note that the field $\mathbf{E}_{\omega, dipole}$ is responsible for the molecule-nanocrystal near-field interaction. Then, the total absorption in our system is composed of two terms:

$$Q = Q_{molecule} + Q_{NC}.$$

Equation 3 and 4 give the absorptions of the components.

## 2.2. Circular dichroism

The CD cross section should be calculated as the difference:

$$CD = \langle Q_+ - Q_- \rangle_\Omega, \tag{5}$$

where $\langle ... \rangle_\Omega$ denotes the averaging of light incidence over a solid angle. This averaging is needed because these hybrid complexes are in a solution and have random orientations. The rates $Q_{+(-)}$ represent the total absorptions by the complexes for the left and right circularly polarized beams. It is convenient to split the total CD signal into two parts:

$$CD = CD_{molecule} + CD_{NC},$$

where the terms are given by

$$CD_{molecule} = \langle Q_{molecule,+} - Q_{molecule,-} \rangle_{\Omega},$$

$$CD_{NC} = \langle Q_{NC,+} - Q_{NC,-} \rangle_{\Omega}.$$

These terms come from the absorption of biomolecule and NCs, respectively. We note that, although the total dissipation rate is split mathematically into two parts, the molecular and plasmonic components are strongly interacting. According to refs. [5,8] the CD rates should be calculated from the following equations:

$$CD_{molecule} = E_0^2 \frac{8}{3}\sqrt{\varepsilon_0}\omega_0 \frac{\Gamma_{12}}{|\hbar\omega - \hbar\omega_0 + i\Gamma_{12}|^2} \text{Im}\left[\left(\hat{\mathbf{P}}^{(tr)*} \cdot \boldsymbol{\mu}_{12}\right) \cdot \mathbf{m}_{21}\right],$$

$$CD_{NC-molecule} = \text{Im}(\varepsilon_{NP})\frac{\omega}{2\pi} \cdot E_0^2 \frac{4\sqrt{\varepsilon_0}}{3} \cdot \text{Im}\left[\frac{1}{\hbar\omega - \hbar\omega_0 + i\Gamma_{12}} \int_{metal} \left(\hat{\mathbf{K}}^{(tr)*} \cdot \vec{\nabla}\left(\Phi_{\omega} \cdot \boldsymbol{\mu}_{12}\right)\right) \cdot \mathbf{m}_{21} \cdot dV\right]$$

(6)

where $\boldsymbol{\mu}_{12}$ and $\mathbf{m}_{21}$ are the matrix elements of the involved quantum states $|1\rangle$ and $|2\rangle$; $\varepsilon_0$ is the dielectric function of the liquid matrix, and we use $\varepsilon_0 = 1.8$ (water). The matrices $\hat{\mathbf{P}}$ and $\hat{\mathbf{K}}$ describe the fields at the molecule and inside the NPs. Specifically, the matrix $\hat{\mathbf{P}}$ determine the total field at the molecule:

$$\mathbf{E}^{(0)}_{\omega,ind}(\mathbf{R}) = E_0 \left( \hat{\mathbf{P}} \cdot \mathbf{e}_0 \right), \tag{7}$$

where $\mathbf{e}_0$ is the polarization of the external electromagnetic wave; in fact, $\hat{\mathbf{P}}$ can be regarded as an enhancement matrix. The second matrix in Equation 6 is given via the following equation:

$$\mathbf{E}^{(0)}_{\omega,ind}(\mathbf{r}) = E_0 \left( \hat{\mathbf{K}}(\mathbf{r}) \cdot \mathbf{e}_0 \right). \tag{8}$$

The matrix $\hat{\mathbf{K}}(\mathbf{r})$ describes the electric fields inside the NCs in the absence of the molecule and is used inside the integral in Equation 6. The position-dependent vector $\mathbf{\Phi}_\omega$ is a response function of the NCs for a point dipole placed at the position of a molecule; it is defined in the following way: $\varphi_{\omega,dipole} = \mathbf{\Phi}_\omega \cdot \mathbf{d}_\omega$, where $\mathbf{d}_\omega$ is a probe dipole and $\varphi_{\omega,dipole}$ is the induced potential. For more details about Equation 6, one can see the derivations given in refs. [4,5,8].

Equations 6 look rather complex, but they do have clear physical meanings. The component $CD_{molecule}$ is the plasmon-enhanced CD spectrum of a molecule. This CD signal appears, of course, at the wavelength of the molecular transition. The term $CD_{NC-molecule}$ is more peculiar – it describes the plasmon resonances of the NCs. We should note that, from the very beginning, the plasmonic system is assumed to be non-chiral and the chiral plasmonic signal, $CD_{NC-molecule}$, occurs only owing to the chiral molecule. The chiral molecular dipole interacts with the NCs via near-field Coulomb fields and induces plasmonic dissipative currents inside the NCs. These dissipation currents lead to absorption of light and, consequently, to the plasmon-induced CD signals.

# 3. Characterization of hot spots of the plasmonic systems

CD signals given by Equation 6 involve induced electric fields inside the system. Therefore, it is instructive to look first at the electromagnetic enhancement factors for the hybrid complexes. For this we calculate the following quantity:

$$P(r = R_{mol}) = \frac{|\mathbf{E}_\omega|^2}{E_0^2}$$

where $R_{mol}$ is the position of the molecule in our system, and $\mathbf{E}_\omega$ and $E_0$ are the actual field at the molecular position and the incident field amplitude, respectively. In **Figure 2** we show only the geometry excited with $\mathbf{E}_0 \parallel \hat{\mathbf{z}}$, in which the plasmonic NPs can create strong enhancement at the position of the biomolecule. As expected, we see strong enhancement effects for this polarization of the incident light. Remarkably, the Al-NCs in Figure 2 exhibit stronger enhancements factors compared to the traditional Au- and Ag-NCs. Moreover, as expected, the Al-NCs have the plasmon peak in the UV spectral region. The NC dimer systems in Figure 2 show stronger enhancement and the spatial gap region in this system can be regarded as an electromagnetic hot spot. Of course, we would like to place the chiral molecule into such hot spot regions.[5] We note that, in the other field configuration ($\mathbf{E}_0 \perp \hat{\mathbf{z}}$), the enhancement factors are typically small because the NCs tend to screen the external field. In Figure 2, we show only the plasmonic NPs with large resonant enhancement factors. The Si- and glass-NCs have relatively small enhancements, with $P \sim 1$.

## 4. Plasmon-induced and plasmon-enhanced CD mechanisms and Fano-like interference effects

### 4.1 Quantum theory of molecular CD

When a light beam propagates through a collection of chiral biomolecules dispersed in a solution, it simultaneously exhibits two effects: Circular Dichroism (CD) and Optical Rotatory Dispersion (ORD) effects. The rotational effect (ORD) and the differential absorption (CD) are intrinsically related. Mathematically, they are connected by Kramers-Kronig relations.[1,8] These phenomena have been intensively investigated and are well understood.[1,2] Within the macroscopic chiral medium approach, the CD and ORD effects are proportional to the so-called chiral parameter, $\xi_c(\omega)$.[38] Specifically,

$$CD \propto \text{Im}[\xi_c], \qquad ORD \propto \text{Re}[\xi_c].$$

A convenient expression for the chiral parameter was derived in ref. [35] and it reads:

$$\xi_c = 4\pi n_0 \frac{i(\boldsymbol{\mu}_{12} \cdot \mathbf{m}_{21})}{3} \left( \frac{1}{\hbar\omega - \hbar\omega_0 + i\Gamma} + \frac{1}{\hbar\omega + \hbar\omega_0 + i\Gamma} \right), \qquad (9)$$

where $\hbar\omega_0$ and $\Gamma = \Gamma_{12}$ are the molecular resonance energy and its broadening, respectively; $\boldsymbol{\mu}_{12}$ and $\mathbf{m}_{21}$ are the matrix elements of the optical dipoles as defined above and $n_0$ is the density of randomly-oriented chiral biomolecules. The product $\boldsymbol{\mu}_{12} \cdot \mathbf{m}_{21}$ in Equation 9 is an imaginary number, i.e. $\boldsymbol{\mu}_{12} \cdot \mathbf{m}_{21} = i \cdot a$, where $a$ is real. Then, near the resonance $\omega \approx \omega_0$, we obtain the spectra, which are usually regarded as the famous Cotton effect:[1]

$$CD_{mol} \propto \frac{-\Gamma}{\left(\hbar\omega - \hbar\omega_0\right)^2 + \Gamma^2}, \qquad ORD_{mol} \propto \frac{\hbar\omega - \hbar\omega_0}{\left(\hbar\omega - \hbar\omega_0\right)^2 + \Gamma^2}.$$

Below we will use these well-established and fundamental properties of optical activity in the understanding of the plasmon-induced and plasmon-enhanced CD mechanisms.

**4.2 Circular dichroism of a hybrid complex composed of one NC and one molecule**

Now we will look at the case of a hybrid structure composed of one molecular element and one plasmonic NC (Figs. 3 and 4). The chiral molecular transition can be in resonance with a plasmonic NC or it can be off resonance. As an example, we will consider the case of a molecule that is not in strong resonance. For our calculations, we chose a molecule with an optical transition at a wavelength of 300 nm. This transition sits at a higher energy than the Au- and Ag-NCs resonances, but lower than the one for the Al plasmon. For the optical dipoles of a biomolecule, we take typical numbers [5]. In particular, we first write the optical dipole moments in the following way: $\mu_{12} = |e| \cdot r_{12}$ and $(\mathbf{\mu}_{12} \cdot \mathbf{m}_{21}) / \mu_{12} = -i \cdot |e| r_0 \cdot r_{12} \cdot \omega_0 / (2c_0)$. Then, we define the constituting parameters: $\lambda_0 = 300\ nm$, $\omega_0 = 4.13\ eV$, $r_{12} = 2\ \text{Å}$, $r_0 = 0.05\ \text{Å}$, and $\Gamma_{12} = 0.3\ eV$. For an isolated chiral molecule, the above parameters yield typical numbers for the extinction and CD cross sections. In Figure 3, we also observe that the Al-NC has the strongest extinction and the molecular transition is, of course, the weakest one. In addition, the Al-NCs, as it was mentioned above, have the strongest enhancement factors.

*4.2.1. Plasmon-induced and plasmon-enhanced CD mechanisms.*

The CD spectra of the complexes have a characteristic structure **(Figure 4)**, with a molecular transition peak and a plasmonic band. The plasmonic band can be either positive or negative, depending on two factors: (1) the orientation of the molecular dipole with respect to the surface and (2) the spectral position of the molecule with respect to the plasmonic band of a NC. We now consider two different cases.

*Case 1: Al-NCs with $\lambda_{plasmon} < \lambda_{mol}$*

The plasmon-induced CD band is negative and strongest for the normal dipolar orientation $\boldsymbol{\mu}_{12} \parallel \hat{z}$. For the case of $\boldsymbol{\mu}_{12} \perp \hat{z}$, the plasmon CD band is positive and relatively weak.

*Case 2: Ag- and Au-NCs with $\lambda_{plasmon} > \lambda_{mol}$*

The behaviours are opposite to that of the previous case. We now observe the positive plasmonic bands in CD spectra for the orientation $\boldsymbol{\mu}_{12} \parallel \hat{z}$. For the orientation $\boldsymbol{\mu}_{12} \perp \hat{z}$, the plasmonic bands are negative. Also, the plasmon-induced signals are much weaker than those in the case of Al-NCs. The latter is expected since the Al-NCs have much larger enhancement factors.

The above observations can be understood by looking at the dipolar limit for the equation 6. This limit was analysed in detail in refs. [4,8]. If we reduce the molecule-NC interaction to the dipole limit, we see that

$$CD_{NC, dipole-external\ field} \propto F_{res}(\omega) \cdot \frac{a_{NP}^3}{L^3} \cdot \text{Im}(\varepsilon_{NP}) \cdot \text{Im}\left[\mu_{12x} \cdot m_{21x} + \mu_{12y} \cdot m_{21y} - 2 \cdot \mu_{12z} \cdot m_{21z}\right] \frac{\hbar\omega - \hbar\omega_0}{(\hbar\omega - \hbar\omega_0)^2 + \Gamma^2}.$$

Here $F_{res}(\omega)$ is the resonant factor, and $a_{NP}$ and $L$ are the NP size and the molecule-NP center-to-center separation, respectively. This equation reproduces all above behaviours for the signs of the plasmon-induced CD bands. We also see one fundamental property of the plasmon-induced CD:

$$CD_{NC,dipole-external\ field} \propto F_{res}(\omega) \cdot \frac{a_{NP}^3}{L^3} \cdot \text{Im}(\varepsilon_{NP}) \cdot ORD_{mol}.$$

This equation tells us that the origins of the plasmon-induced CD are in: (1) the ORD of the chiral molecule and (2) the energy dissipation inside a plasmonic NC. Simultaneously, we see that the derived CD signal also contains the plasmon resonant factor $F_{res}(\omega)$ and is therefore enhanced at the plasmon resonance.

In Figure 4, we also show for comparison the data for semiconductor and dielectric NCs with small sizes. We observe that such dielectric and absorbing objects do not create strong effects in the CD spectra. There are two reasons for this: (1) field enhancement factors are not large because they have no resonances and (2) the near-field interactions are not strong. In the spectrum of the Si-based complex, one can see some Fano-type interference effect that comes from the absorption in the Si-NC (Figure 4b). However, when we involve Si-NCs with larger sizes, we can expect interesting optical phenomena in the CD spectra. Large dielectric nanocrystals with sizes comparable to the wavelength of light form optical resonators with a rich family of Mie resonances which may result in interesting effects in CD spectra.[39]

## 5. Plasmonic hot spots: Plasmon-induced CD in nanocrystal dimers with a chiral biomolecule

We now look at the most interesting case, that of a nanostructure with a plasmonic hot spot. In general, the plasmonic CD effects are expected to be strongly enhanced in the NP dimers, as compared to the single-NP complexes, because a chiral biomolecule is now being placed in the so-called hot-spot region.[5,9] First, we consider the case of the Al-Al dimer, with the strongest hot spot **(Figure 5)**. The strongest CD signals are again found for the configuration with the molecule dipole perpendicular to the NC surfaces (Figure 5). This is the configuration with $\mu_{12} \parallel \hat{z}$. The other configuration, with $\mu_{12} \parallel \hat{x}$, has again smaller CD signals. This is expected, since plasmonic field-enhancements factors are much larger for excitation along the long symmetry axis of the system. The patterns and the signs of the CD peaks in Figure 5, which describe the Al-Al complex, are more complex and somewhat different to the spectra of a single Al-NC (Figure 4a). The reason is that, in the Al-Al dimer, many multipolar harmonics contribute to the near-field, whereas the spectra for a single Al-NC involves mostly the dipolar near-field. In the plasmonic dimer structure, the CD spectrum becomes more complex and we can qualitatively summarize it in the following equation:

$$CD_{NC,dipole-external\ field} \propto g_{res}(\omega) \cdot f(\mu_{12}, m_{21}) \cdot \mathrm{Im}(\varepsilon_{NP}) \cdot \frac{\hbar\omega - \hbar\omega_0}{(\hbar\omega - \hbar\omega_0)^2 + \Gamma^2}, \qquad (10)$$

where $g_{res}(\omega)$ and $f(\mu_{12}, m_{21})$ are complex functions which are not known analytically; $g_{res}(\omega)$ describes the resonant behaviour in the hot spot in terms of the multipolar plasmons of the dimer, and $f(\mu_{12}, m_{21})$ reflects the origin of the chiral signal in the biomolecule. In the dipolar limit of the NC-NC interaction, Equation 10 is given in ref. [8].

Using the same formalism, we also computed the Ag-Ag, Au-Au, Al-Ag and Al-Au systems. In the Ag-Ag and Au-Au complexes, we typically observe a more complex structure of the plasmonic CD band **(Figure 6)**. This is again owing to the involvement of the multipole resonances.[5,9] In the hybrid, heterogeneous complexes (Al-Ag and Al-Au, **Figure 7**), we found two plasmonic bands related to the two metals involved in the complex. In these cases the CD bands of the two plasmons in the complex alternate in sign, owing to the molecule-NC interactions with two different materials.

**5.1. The sensitivity of the plasmonic CD**

It is interesting to point out that, like CD spectroscopy of biomolecules alone, plasmonic CD is a very sensitive method of detection. Typical cross sections of NCs are much greater than those for molecules (Figure 3). Therefore, the molecular contribution to the total extinction cross section of a molecule-NP complex is tiny and the contribution of the metal NPs governs the extinction of the complex (Figure 3b). Nevertheless, the CD spectrum of a hybrid complex demonstrates strong CD effects owing to both the chiral molecule and the molecular-NP interaction. The reason why we see the molecule CD effect in the CD spectrum of the complex is that the plasmonic NC system itself is not chiral, so chirality in the system comes from the biomolecular component. Simultaneously, the role of the plasmonic NCs is twofold: (1) they create new CD bands as compared to the original molecular systems and (2) enhance the CD signals.

Regarding the sensitivity of the plasmon-induced CD approach, the strength of CD signals in a complex really depends on the design and composition. In particular, we see that the plasmon-induced CD signals decrease dramatically with increasing molecule-NC distances (Figure 4 and 5). The other crucial factors are: (1) the molecular orientation, (2) the spectral difference between the molecular wavelength and the plasmon resonance and (3) the strength and broadening of the plasmon resonance used in the complex. Using the above properties, we can understand why the Al-based complexes give the strongest CD signals as compared to the ones based on Au and Ag.

6. Discussion and conclusions

Finally, we mention some other prominent mechanisms affecting plasmonic CD in NC bio-assemblies and meta-structures. The other sources of the CD effect in plasmonic bio-assemblies come from the plasmon-plasmon interactions. A bio-assembly may have several NCs arranged in a chiral geometry, with the resulting CD effect coming in this case from the NP-NP interactions.[4,40-42] Considering just a single nanocrystal, the CD effect can appear due to a chiral shape.[43,44] In hybrid planar metamaterials structures, chiro-optical effects are generated via long-range electromagnetic interactions between plasmonic modes and a chiral molecular medium.[45,46] Interesting interference between the plasmonic CD effect and intrinsic chirality of biomolecular linkers can be observed in DNA-assembled helices.[47]

      The field of UV plasmonics is presently emerging. It is really interesting to use metals with UV plasmons for enhancing molecular and bio-molecular signals. We know that optical properties can be very different when exciting materials in the visible and UV intervals. In that sense, investigations of plasmon-enhanced effects, such as SERS and CD, in the UV spectral interval is something that can bring new and surprising results. Aluminum was naturally identified as one of the most promising metals for UV plasmonics [48]. In the available literature, this metal was mostly considered in the context of the SERS effect. This theoretical study described the CD effect induced by the Al UV-plasmons and identified geometries and

parameters of Al nanoparticles that can give significant enhancement and also lead to new CD spectral lines. One possible complication of UV plasmonics with aluminum is the oxide layer [24,25] that can shift the plasmon resonance to the red and also impose a limitation for the NC-NC gap. The calculations presented here were done for the simplest model of Al-Al dimers without the oxide layers. Our goal was to show the potential of aluminum for the chiral biomolecular studies, a goal for which it is indeed a unique and promising material. However, the influence of the oxide layer is an important issue and it should be further investigated. We should also note that, along with aluminum, the current literature on UV plasmonics considers several materials [48] such as chromium, copper, gallium, indium, magnesium [49], palladium, etc.

In this study, we considered the case of a non-chiral plasmonic nanostructure interacting with a chiral molecule. To date, such hybrids have been reported by many groups.[3,6] The focus of this paper was the possibility of using aluminum as an UV-plasmonic element. Indeed, we see that Al-NCs can be advantageous for the amplification of CD signals of biomolecules for two reasons: (1) the plasmon resonances in the Al NCs are close to the UV bands of typical biomolecules and (2) the Al material system can provide us with much larger enhancement factors. In real chiral biomolecules, one can typically observe strong CD lines in the UV range, where we also predict the plasmon-induced CD signals. However, we can differentiate the natural CD lines and the plasmon-induced peaks since: (1) the plasmonic resonances can be stronger and (2) the plasmonic lines have well-defined wavelengths. Finally, we note that potential applications of chiral nanocrystals,[50,51] including plasmonic ones, exist in fields that can take advantage of the sensing and recognition of biomolecules, such as cell studies [52] and drug development.

**ACKNOWLEDGMENTS**

This work was supported by Volkswagen Foundation (Germany) and by the Army Office of Research (MURI Grant W911NF-12-1-0407). A.O.G. acknowledges support via Chang Jiang Chair Professorship (China).

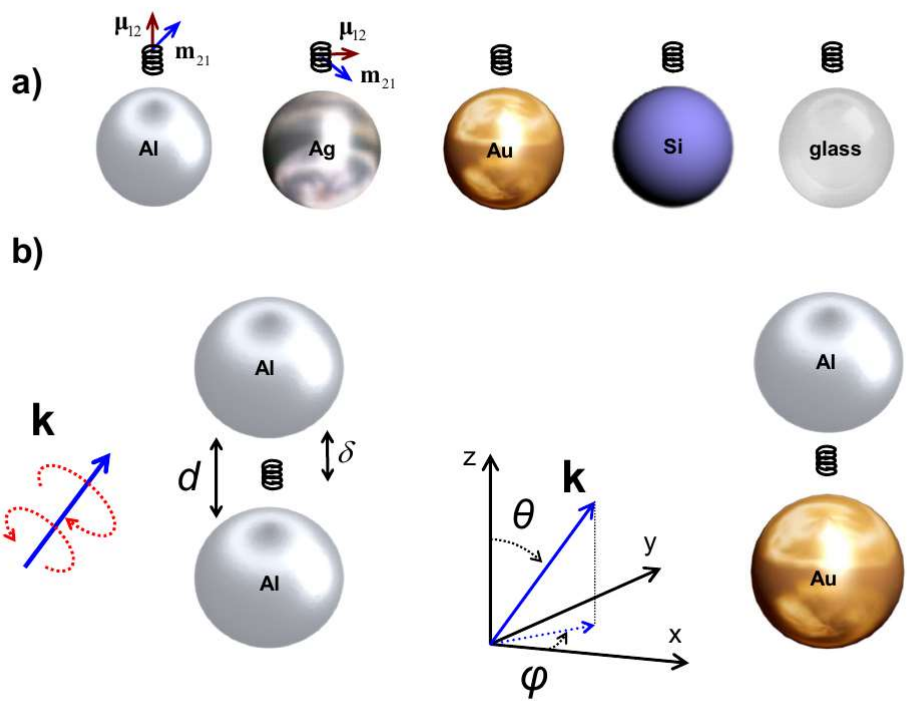

**Figure 1:** a) Models of hybrid complexes composed of a chiral molecule and a single crystalline NC. These models involve metal, semiconductor and dielectric materials. **b)** Models of NC dimers with a chiral molecule placed in the hot-spot region.

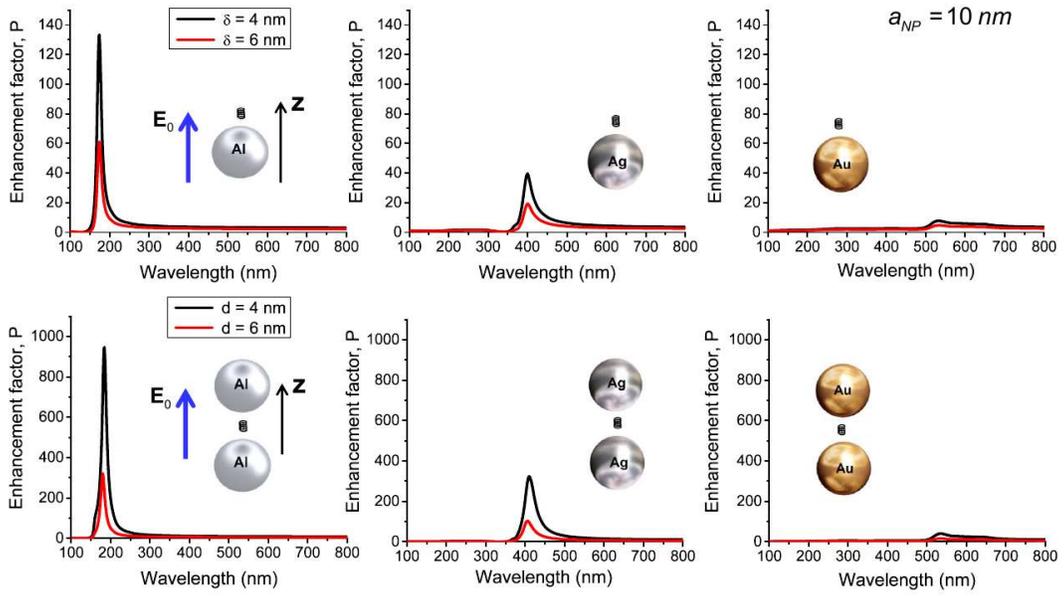

**Figure 2:** Electromagnetic enhancement factors of a chiral molecule for the plasmonic NC complexes. All NCs have the same diameter, $a_{NP} = 10\,nm$. In our calculations, we use a molecule with a resonance at $\lambda = 300\,nm$. Inserts show the geometries.

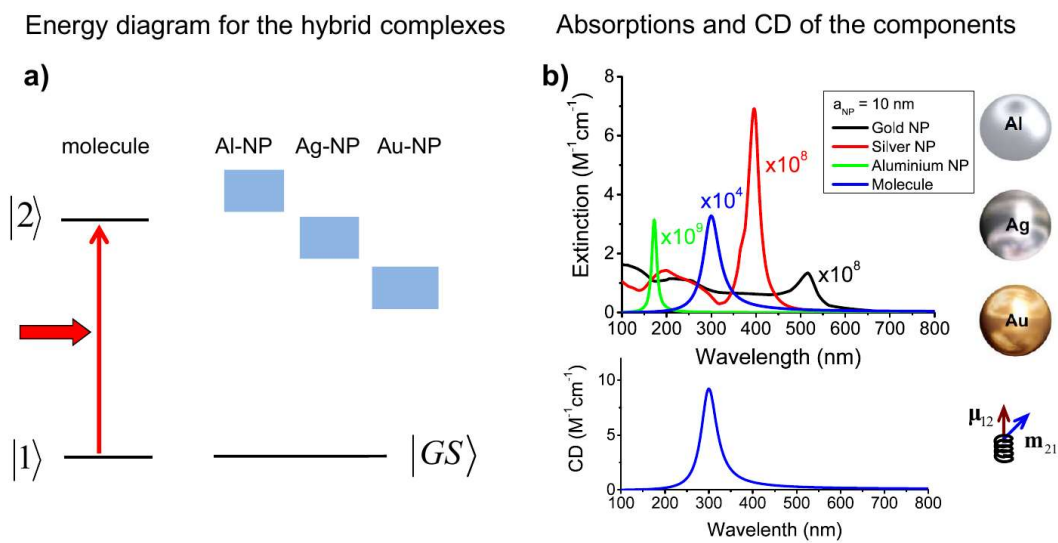

**Figure 3: a)** Energy diagram for the molecular and plasmonic energies in the hybrid complexes. This diagram shows the discrete molecular transition and the three different plasmonic bands of the NCs. **b)** Extinctions and CD spectra of the molecular and plasmonic components. Insets: Models.

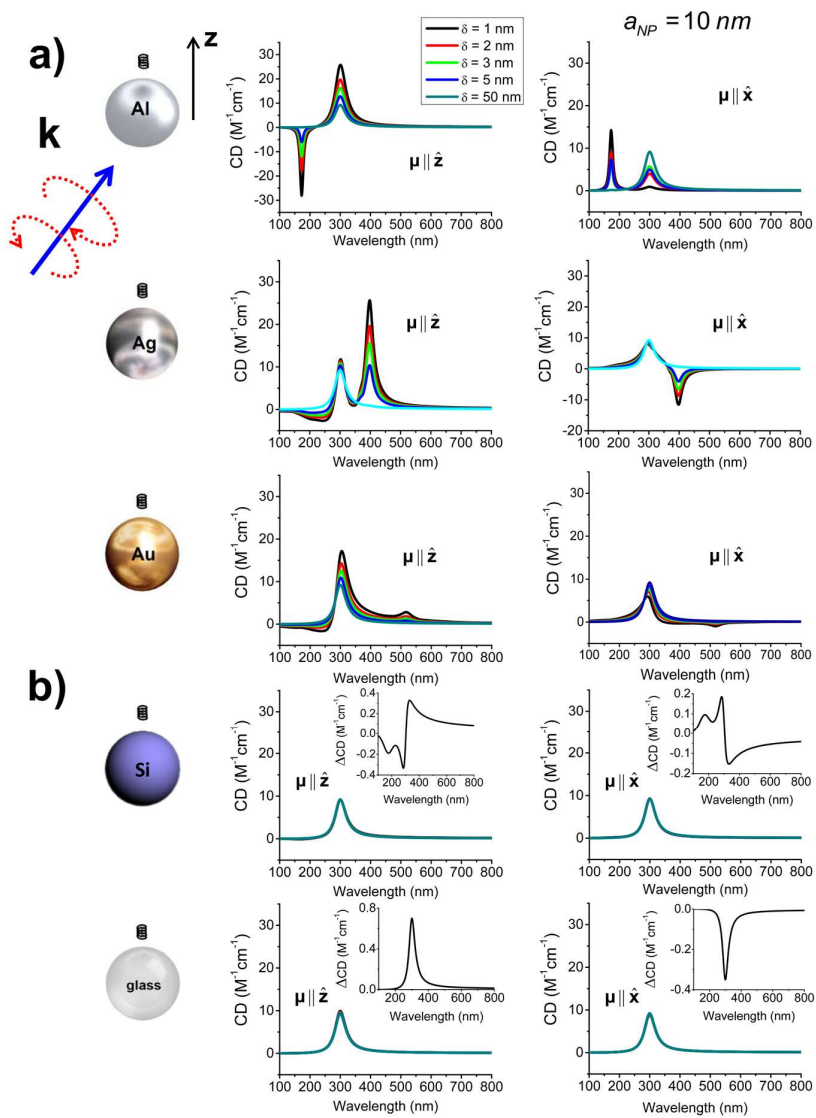

**Figure 4:** Calculated CD spectra of the hybrid complexes incorporating one crystalline nanoparticle and one chiral molecule. Panels **(a)** show CD spectra of plasmonic NPs and panels **(b)** present the cases of semiconductor and dielectric NPs. Inserts in figures show the change in the total CD signal in the $\delta = 1\,nm$ system with respect to an isolated molecule. Inserts: Models.

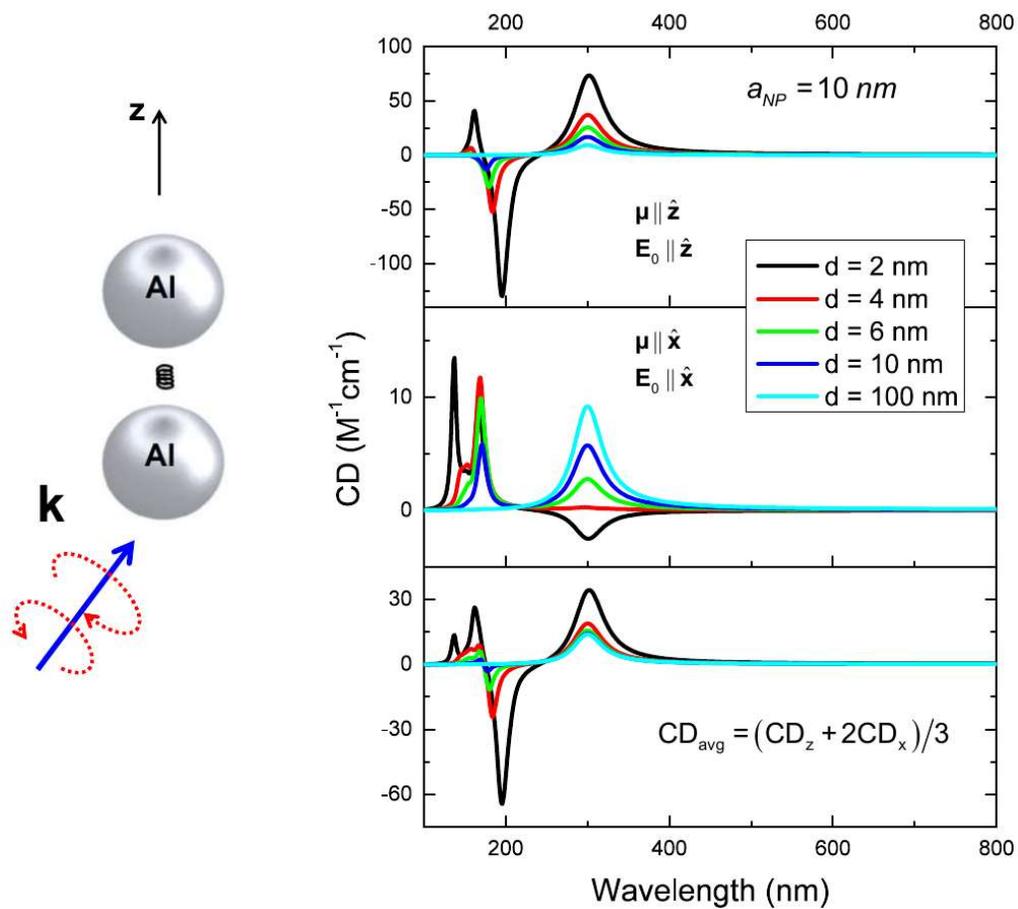

**Figure 5:** Calculated CD spectra of the hybrid complexes incorporating two Al-NCs and one chiral molecule. Molecules are placed in the plasmonic hot spots. Inserts: Models.

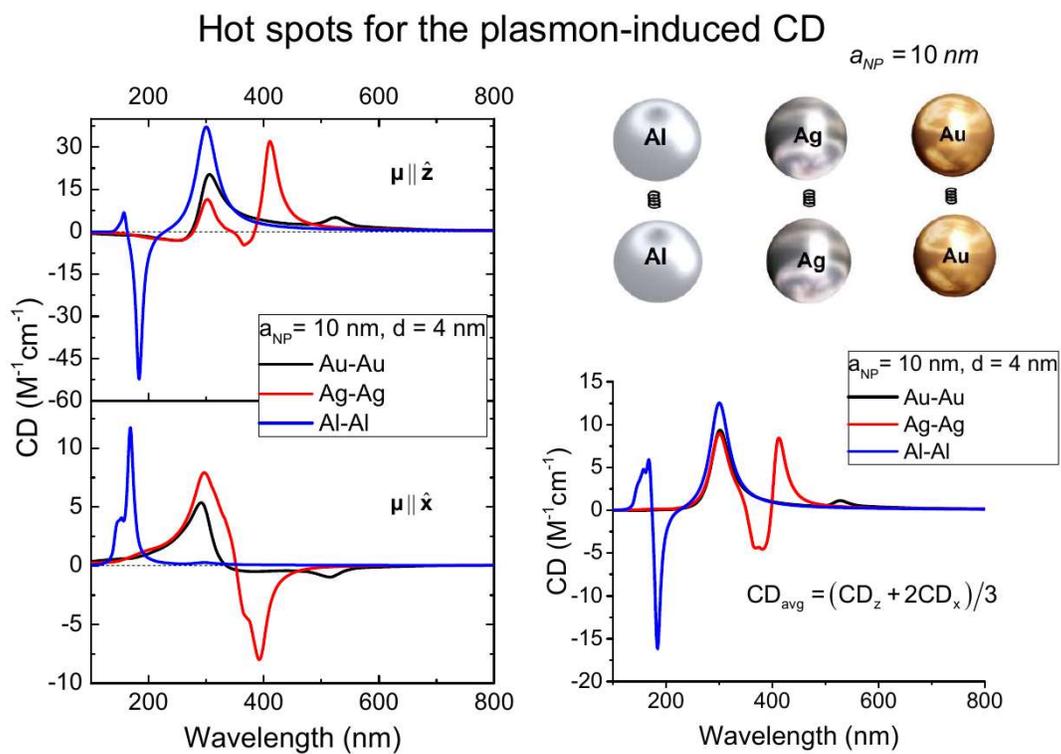

**Figure 6:** Comparison of the Al-Al nanoantenna with the Ag-Ag and Au-Au plasmonic dimers. Again, chiral molecules are placed in the plasmonic hot spots. We also show the CD spectrum for a complex in which the molecule is randomly oriented. Inserts: Models.

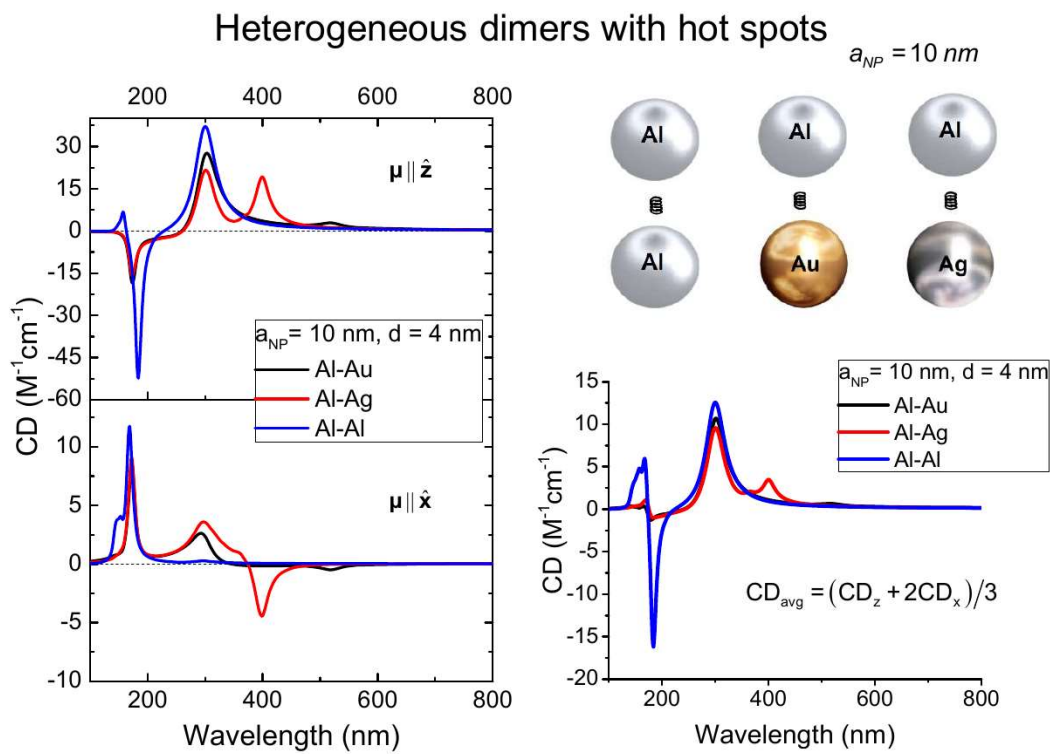

**Figure 7:** CD spectra of the chiral assemblies based on the hybrid dimers: Al-Al, Al-Au, Al-Ag. The CD spectra reveal plasmon peaks of the metal components. The sign of the plasmon-induced CD depends on the relative position of the molecular and plasmonic resonances in each dimer. The graph on the right-hand side shows the CD spectrum with a randomly-oriented molecule in the hot spot. Inserts: Models.